\documentclass[10pt,journal,compsoc]{IEEEtran}
\usepackage{xurl}
\usepackage{color}

\ifCLASSOPTIONcompsoc
  \usepackage[nocompress]{cite}
\else
  \usepackage{cite}
\fi
\ifCLASSINFOpdf
\else
\fi
\begin{document}
%
\title{The Rise of Quantum Internet Computing}
%
%
%
%

\author{Seng W. Loke,~\IEEEmembership{Member,~IEEE}
\IEEEcompsocitemizethanks{\IEEEcompsocthanksitem Seng W. Loke is with the School of Information Technology, Deakin University, Melbourne, Australia.\protect\\
E-mail: see https://www.deakin.edu.au/about-deakin/people/seng-loke.}
\thanks{Manuscript received X XX, 20XX; revised X XX, 20XX.}}

%
%

\markboth{IEEE IoT Magazine,~Vol.~XX, No.~X, X~2022}%
{Loke \MakeLowercase{\textit{et al.}}: Towards Quantum Internet Computing}
%



\IEEEtitleabstractindextext{%
\begin{abstract}
This article highlights {\em quantum Internet computing} as referring to  distributed quantum computing over the quantum Internet, analogous to (classical) Internet computing involving (classical) distributed computing over the (classical) Internet. Relevant to quantum Internet computing would be areas of study such as quantum protocols for distributed nodes using quantum information for computations, quantum cloud computing, delegated verifiable blind or private computing,  non-local gates, and distributed quantum applications, over Internet-scale distances.

\end{abstract}

\begin{IEEEkeywords}
quantum Internet computing, quantum Internet, distributed quantum computing, Internet computing, distributed systems, Internet\\
{\color{red}
"This work has been submitted to the IEEE for possible publication. Copyright may be transferred without notice, after which this version may no longer be accessible."
}
\end{IEEEkeywords}}

\maketitle

\IEEEdisplaynontitleabstractindextext

%
\IEEEpeerreviewmaketitle

\IEEEraisesectionheading{\section{Introduction}\label{sec:introduction}}

%
%
%
%
\IEEEPARstart{T}{here} have been tremendous developments in quantum computing, quantum cryptography, quantum communications and the quantum Internet, and we have seen   increased investments and intensive research in quantum computing in recent years~\cite{koz19,rohde_2021}. The quantum Internet will not necessarily replace the (classical) Internet we know and use today, at least not in the near future, but can complement the current Internet. The quantum Internet aims to enable robust   {\em quantum teleportation} (or transmission) of qubits,\footnote{A qubit is the basic unit of quantum information, and can be thought of as a two-state, or   two-levelled, quantum-mechanical system, such as an electron's spin, where the two levels are spin up and spin down, or a photon's polarization, where the two states are the vertical polarization and the horizontal polarization.} and {\em entanglement}  among qubits,\footnote{Multiple qubits at different sites can share an entangled state, a  superpositon of ``specially correlated'' states, to be used in distributed algorithms.} over long Internet-scale distances, which are key to many of the quantum protocols including quantum key distribution, quantum voting, and others, as well as for non-local control of quantum gates. 

There have been efforts to build quantum computers, and it remains to see if any one paradigm becomes the dominant or best way of building such quantum computers. At the same time, even as researchers develop more powerful quantum computers (supporting more qubits for operations, and at lower error rates), there is an opportunity for connecting multiple quantum computers from different sites to achieve much more complex quantum computations, i.e., inter-linking multiple quantum computers on different sites to perform distributed computing with a distributed system of quantum computers (or quantum processing units (QPUs) at different nodes), arriving at  the notion of {\em distributed quantum computing}, e.g., \cite{parekh21}.

While distributed quantum computing can involve multiple QPUs next to each other or at the same site, with the quantum Internet, one can envision  distributed quantum computing over nodes geographically far apart. As noted in~\cite{cuomo20}, the idea is the quantum Internet as the ``underlying infrastructure of the Distributed Quantum Computing ecosystem.''

This article highlights the emerging area of   distributed quantum computing over the quantum Internet, which we refer to as {\em quantum Internet computing}, i.e., the idea of computing using quantumly connected distributed quantum computers over Internet-scale distances. Hence, quantum Internet computing is not a new concept in itself but a proposed ``umbrella term'' used here for the collection of topics (listed below), from an analogy to (classical) Internet computing. 

{\em Internet computing}, where one does distributed computing but over Internet-scale distances and distributed systems involve nodes connected via the Internet, is at the intersection of work in (classical) distributed computing and the (classical) Internet. Analogous to Internet computing, one could ask the question of what would be at the intersection of work in {\em distributed quantum computing} and work   on the {\em quantum Internet}, which  brings us to the notion of  {\em quantum Internet computing}.

Also, 
while the quantum Internet and distributed quantum computing are still nascent research areas,  there are at least three key topics which can be considered as relevant to quantum Internet computing:
\begin{itemize}
\item	distributed quantum computing, including quantum protocols from  theoretical perspectives involving communication complexity studies, and
	distributed quantum computing via non-local or distributed quantum gates,
\item	quantum cloud computing with a focus on delegating quantum computations, blind quantum computing, and verifying delegated quantum computations, and
\item  computations and algorithms for the	quantum Internet including  key ideas such as quantum entanglement distillation, entanglement swapping,  quantum repeaters,  and quantum Internet standards.\footnote{For example, see \url{https://www.ietf.org/archive/id/draft-irtf-qirg-principles-10.html} [last accessed: 1/8/2022]} 
\end{itemize}
We briefly discuss the above topics in the following sections.

\section{Distributed Quantum Computing}
Distributed quantum computing problems and quantum protocols have been well-studied for over two decades, from a theoretical computer science perspective,\footnote{For example, see Buhrman and Röhrig's  paper dating  back to 2003: \url{https://link.springer.com/chapter/10.1007/978-3-540-45138-9_1} [last accessed: 1/8/2022]} many of which have their inspiration from classical distributed computing research. Quantum versions of classical distributed computing problems and protocols, and new forms of distributed computing using  quantum information, have been explored, e.g., the distributed three-party product problem, the distributed Deutsch-Jozsa promise problem and the distributed intersection problem, demonstrating how, for some problems, quantum information can enable fewer bits of communication to be used for a solution, and  how certain distributed computation problems can be solved with quantum information, but cannot be solved  classically. Many  quantum protocols, including quantum coin flipping, quantum leader election,   quantum anonymous broadcasting, quantum voting, quantum Byzantine Generals, quantum secret sharing, and quantum oblivious transfer, can be viewed as ``quantum versions'' of classical distributed computing problems, and have been studied extensively.

Another area of study, which has also been considered as distributed quantum computing, is non-local gates, or the non-local control of quantum gates, including early work nearly over two decades ago.\footnote{For example, see the work by Yimsiriwattana and Lomonaco Jr. in  \url{https://arxiv.org/pdf/quant-ph/0402148.pdf} and a distributed version of Shor's famous factorization algorithm  \url{https://arxiv.org/abs/2207.05976} [last accessed: 1/8/2022]} Key to performing such non-local control of quantum gates is the use of entanglement, which can be viewed as a resource for such non-local computations. More recent work has looked at how to partition  the computations of distributed quantum circuits  over multiple QPUs, e.g., \cite{parekh21} as we mentioned earlier - with considerations including  distributing computations in such a way as to optimize performance and to  reduce the requirements on entanglement, since if the entanglements required are generated at too low a rate, this will hold up computations.  The key motivation here is to inter-link a set of quantum computers to form effectively a much more powerful quantum computer.

\section{Quantum Cloud Computing and Delegating Quantum Computations}
We have seen big tech companies and startups offering quantum computing as a service similar to accessing other cloud service offerings, which is a fantastic resource for experimentation and studies. 

More formally, studies into delegating quantum computation from a client (which can be either classical, or almost classical, i.e., with minimal capability to perform computations such as  teleporting qubits, applying simple Pauli quantum operations, and doing basic  measurements) which is much more restricted than the server (assumed to be a universal quantum computer) have been studied, called {\em delegated quantum computing}. And when the server is prevented from knowing the client's inputs but still can perform delegated computations, by a technique such as the {\em quantum one-time pad} (where the client applies  Pauli operations to add uncertainty from the server's perspective,  thereby effectively encrypting the  quantum inputs it sends to the server, and keeps track of operations it later needs to decrypt the outputs from the server), this is called {\em blind} quantum computing. 

In order to be sure that the server does indeed perform the required quantum operations delegated to it by the client,   the client can embed tests (or test runs) into the delegated computations, so that the server (not being able to distinguish between tests and normal computations) can be caught out if it did not perform the required computations properly. That is, the client can verify if the server performed the required quantum computations.\footnote{An excellent example is the work by Broadbent at \url{https://theoryofcomputing.org/articles/v014a011/} [last accessed: 1/8/2022]} 
Further abstractions for delegating quantum computations with supporting cloud services continues to be investigated.

\section{The Quantum Internet}
As we mentioned earlier, work on the quantum Internet focuses on how to efficiently enable robust entanglement shared among qubits over long geographical distances. If two nodes in different continents share entangled states, then, this can be a resource to do non-local gates, i.e., to perform distributed quantum computations, and enable quantum protocols over Internet-scale distances. 

There have been the use of satellites to enable long distance entanglement, as well as the use of optical fibre cables to demonstrate entanglement. Key to the quantum Internet are ideas such as entanglement swapping and quantum repeaters, including ideas such  quantum distillation, to achieve high fidelity distributed entangled states over long distances,  and quantum error correction  - this continues to be a research endeavour as mentioned earlier~\cite{rohde_2021}.

There are other interesting distributed quantum applications to be considered including quantum cryptography, 
quantum sensing, and quantum positioning systems.

\section{Distributed Quantum Computing over the Quantum Internet: Quantum Internet Computing and the Quantum IoT?}

Apart from the many quantum computers available over the cloud by big tech and startups which work at very low temperatures, room temperature quantum computers have also started to emerge.\footnote{See \url{https://spectrum.ieee.org/nitrogen-vacancy-diamond-quantum-computer-accelerator-qubits-server-rack} and also \url{https://otd.harvard.edu/explore-innovation/technologies/scalable-room-temperature-solid-state-quantum-information-processor/}  [last accessed: 1/8/2022]} This could pave the way for quantum computers at the fog and at the edge, not just in remote clouds, and perhaps even mobile quantum computers, or quantum computers embedded into everyday devices and objects, if ever! Will we then have the quantum Internet of Things (IoT)? The answer remains to be seen, and ``quantum entangled things across the world'' will likely complement the classical IoT. Future applications and potential of  quantum Internet computing remains to be investigated. Meanwhile, others have begun to look at the connection between 6G networking and the quantum Internet~\cite{9749227}.

\bibliographystyle{IEEEtran}
\bibliography{qrefs} 
\end{document}